\documentclass[a4paper,12pt]{article}
\usepackage{hyperref}
\usepackage[english]{babel}
\usepackage[T1]{fontenc}
\usepackage[utf8]{inputenc}
\usepackage{amsfonts}
\usepackage{amssymb}
\usepackage{amsmath}
\usepackage{graphicx}
\usepackage{cite}
\usepackage{epsfig}
\usepackage{psfrag}
\usepackage{tikz}
\usetikzlibrary{matrix,arrows,decorations.pathmorphing,positioning} 
\usepackage{url}
\usepackage{amsthm}
\usepackage{mathpartir}
\usepackage{enumerate}
\usepackage{tikz-cd}

\newtheorem{proposition}{Proposition}[section]

\newtheorem{remark}{Remark}[section]

\date{}

\title{\bf  Qualia and the Formal Structure of Meaning }

\author{Xerxes D. Arsiwalla$^{1, 2, }$\footnote{\url{x.d.arsiwalla@gmail.com}}  \\  
{}  \\
{\it \small $^{1}$Pompeu Fabra University, Barcelona, Spain}\\ 
{\it \small $^{2}$Wolfram Institute, Illinois, USA}    
}

\begin{document}

\maketitle

\begin{abstract}

This work explores the hypothesis that subjectively attributed meaning constitutes the  phenomenal content of conscious experience. That is,  phenomenal content is semantic. This form of subjective meaning manifests as an  intrinsic and non-representational character of qualia. Empirically,   subjective meaning is ubiquitous in conscious experiences. We point to phenomenological studies that lend evidence to support this. Furthermore, this notion of meaning closely relates to what Frege refers to as "sense", in  metaphysics and  philosophy of language. It also aligns with Peirce's "interpretant", in semiotics.  We discuss how Frege's sense can also be extended to the raw feels of consciousness.  Sense and reference both play a role in phenomenal experience. Moreover, within the context of the mind-matter relation, we provide a formalization of subjective meaning associated to one's mental representations. Identifying the precise maps between the physical and  mental domains, we argue that syntactic and semantic structures transcend language, and are realized within each of these domains. Formally, meaning is a relational attribute, realized via a map that interprets syntactic structures of a formal system within an appropriate semantic space. The image of this map within the mental domain is what is relevant for experience, and thus comprises  the phenomenal content of qualia. We conclude with possible implications this may have for experience-based theories of consciousness.  

\end{abstract}

\vspace{2pc}

\clearpage

\tableofcontents

\section{Introduction}

How should one formulate a mathematical or metaphysical framework that admits a formalization of the phenomenal contents of conscious experience?  The phenomenal contents or the  so-called qualia of consciousness  are inherently subjective  attributes of experience,  and by most accounts, deemed non-representational  \cite{sep-qualia}.  They refer to  the  raw, immediate and phenomenal aspects of perception, sensation or mental states.  Phenomenal content, by its very nature, is subjective; that is,  available only through one's first-person perspective. Qualia are at the heart of the "what it's like to be" character of conscious experience, as elucidated by Thomas Nagel in 1974 \cite{nagel1974like}. If it were possible to provide a complete third-person description of phenomenal states, that would resolve the "what it's like to be" problem of consciousness.  Qualia  cannot be fully described, communicated or understood through any objective third-person account or computational routine. For instance, attempts to fully convey the subjective experience of tasting chocolate or perceiving the color red through objective descriptions or computational algorithms fall short of capturing the full qualitative nature of those experiences. These observations collectively indicate that certain aspects of conscious experiences pose challenges for all attempts that seek to fully explain or replicate qualia through either exclusively computational, or objective descriptions, that is, potential third-person descriptions. Besides potential  implications for   philosophy of mind, this problem also suggests new frontiers for the philosophy of science itself; namely, how does one extend the existing scientific paradigm based on third-person perspective to account for phenomena that are subjective and non-representational - that is,  amenable only via first-person perspective?  

For these very reasons, a scientific description or theory elucidating the nature of qualia, or one explaining the phenomenology of conscious experiences continues to be an extremely challenging problem at the heart of modern science. Attempts at any such theoretical description (of qualia) call for at least two essential ingredients:  \\  \\  
{\sl (i) A whole new scientific paradigm of subjectivity, one which describes phenomena from the intrinsic point-of-view, or "the view from the inside". \\  \\  (ii) A metaphysical (and possibly  mathematical)  formalization  of non-representational entities (potentially in the form of abstract objects and their  relations).} 

Currently, in the absence of a complete description of (or difficulty conceptualizing) the above two   ingredients, one often relies on constructs using  information, computation or causality \cite{ tononi2004,  dehaene2017consciousness,  oizumi2014phenomenology }. These are certainly  useful to quantify computational processes or identify states of complexity, but that is   still  rather  far from  capturing the very  phenomenal character of qualia\footnote{Roger Penrose has been a vociferous proponent of the view that consciousness necessarily involves a fundamentally non-computational  process  \cite{penrose1994shadows}  (see also \cite{arsiwalla2018brains} for a recent perspective on this point).  An early  remnant of this view, though not explicitly expressed in terms of consciousness,  may be traced back to Alan Turing's PhD thesis, where he contrasts between    intuition and ingenuity  \cite{turing1939systems}. }.  Possibly, a  metaphysical  formalization of phenomenal properties is what is needed. Even though a satisfactory solution to this problem does not exist yet, here we put forth a  proposal that identifies   subjectively attributed meaning  with  the phenomenal content of conscious experience. We then discuss how subjective meaning of this kind can be formalized within the setting of the mind-matter relation.

More broadly, within the cognitive sciences (inclusive of approaches focussing on cognition, consciousness, computation, and clinical practice), meaning has been discussed in one of two ways: either as a by-product of explicit neuronal or computational mechanisms   \cite{dennett1993consciousness},  or as an  intentionality relation between mental representations and what they represent
 \cite{metzinger2004being, atmanspacher2022dual}. These two approaches differ in metaphysical ontology. The former assumes a purely physicalistic stance, whereas the latter presupposes a fundamental relationality  between the physical and the mental.  
 
For instance, in the former scenario,  the meaning associated to one's  experience is believed to arise from various cognitive processes and associations thereof   
 \cite{dennett1993consciousness, johnson2007meaning, galetzka2017story}.   Such cognitive processes may then serve as accessories to conscious experience.  In such a  view, when an individual experiences a scene, it triggers associations, memories, emotions and cognitive mechanisms that provide cognitive significance to that particular experience.  For instance, the color red might evoke emotions, cultural associations or memories linked to the color.  Here,   meaning is not considered an inherent feature of qualia, but something that  is constructed and attributed by an individual based upon their cognitive deliberations and  psychological states. The significance and interpretation of any  experience, is then shaped by one's perceptions, memories, beliefs and also  cultural influences. In other words, meaning construction is discussed  through the lens of cognitive (and metacognitive) mechanistic explanations  \cite{dennett1993consciousness}; that is, as an interpretative layer imposed on top of experience, rather than an inherent feature of raw subjective experience itself.   

In contrast to meaning as a cognitive by-product, the second approach mentioned above,  decrees a more prominent role for meaning with regard to consciousness.  Here, meaning is discussed in the sense of  Brentano's notion of  "intentionality"   \cite{ brentanopsychology }.   In the present work, our discussions align more with  the stance that meaning is indeed fundamental to consciousness. Furthermore, we propose that subjective meaning constitutes the phenomenal content of all conscious experience.  Compiling evidence from phenomenological studies, we demonstrate that this form of  meaning is ubiquitous to the qualia of consciousness.   As mentioned, it is also important to formalize the mathematical or metaphysical  structure of this kind of  meaning,  if one is to make progress on the problem of qualia.  Fortunately,  seminal    works in the philosophy of mind and language  \cite{frege1892sinn } (see  also   \cite{Dummett1983, sep-frege}), as well as mathematical advances in natural language semantics  \cite{coecke2010mathematical}  equip us with the right conceptual groundwork for  formalizing  subjective  meaning with regard to the problem of consciousness. 

The outline of this paper is as follows: in section 2, we provide phenomenological evidence to support the claim that meaning is omnipresent in qualia; in section 3, we discuss how Frege's notion of sense extends to the raw feels of conscious experience; in section 4, we formalize subjective meaning of mental representations in the context of the mind-matter relation; lastly, we conclude in section 5 with final thoughts.


\section{Subjective Meaning in Phenomenology of Experience}

\vspace{.25cm}

\fbox{%
    \parbox{\textwidth}{%
      
\begin{proposition}
  We hypothesize a notion of subjectively attributed meaning or subjective meaning as an inherent feature of the qualia of consciousness. This subjective meaning constitutes the phenomenal content of all conscious experiences. 
\end{proposition}

    }%
}

\vspace{.5cm}


Meaning is not merely about language\footnote{ It was David Mumford who first pointed out that "Grammar isn't merely part of language"  \cite{mumfordblog}. }, nor is it merely a construct about the world "as is".  The notion of meaning has been extensively discussed by several philosophers in the context of language, mind, metaphysics and epistemology    \cite{frege1892sinn, husserl1913ideas, peirce1974collected,  key2017analysis}.  How cognitive and artificial agents learn to make associations about things in the world, is deeply relevant to our understanding of biological and artificial intelligence. 

Discussions about meaning-making or meaning-generation broadly  involve some kind of  attribution of either significance, interpretation or semantic content to an agent's actions, perceptions and thoughts      \cite{dennett1993consciousness, johnson2007meaning, galetzka2017story}.  Some of these attributions may be representable using computational principles, and hence be realized as specific cognitive functions. Alternatively, there may also exist  attributions that are non-representational. These, one might refer to as  phenomenal, and hence intrinsic to the qualia of consciousness.   \\    \\
 {\sf Note that it is this latter class, of non-representational significance attributions, that we will refer to as subjectively attributed meaning or subjective meaning.  }  \\   \\
Let us briefly compare the above two kinds of significance attributions.    Computationally representable attributions are the ones that are particularly well-aligned with an embodied cognition perspective, where an agent has to learn the affordances of its environment for planning its actions and directing itself towards its goals    \cite{shapiro2019embodied,  moulin2017embodied  }.  By some accounts, this process also requires the agent to actively infer or at least conceptualize its own actions and perceptions, as well as that of other agents in the world    \cite{mcdowell1996mind, friston2011action }.  Meaning (in this context of cognitive functions)  pertains to the knowledge or significance that an experience acquires via mechanistic controls or computational processes, over and above its raw phenomenal content. In that sense, this is a deliberative meaning-making process.  For example, meaning associated to  perceptual processes  involve the interpretation and understanding (often deliberative)  of sensory stimuli, such as recognizing various representations of shapes, colors and patterns, or perceiving objects in the external world. In language and cognitive semiotics,  meaning involves the understanding of language-based representations, concepts and symbols, when cognitive agents engage in understanding and communicating through language, signs and gestures  \cite{zlatev2015cognitive}.  
Similarly,  cognitive states involving beliefs, biases or intentions,  involve some kind of value attribution by the individual (based on psychological conditioning)    \cite{song2009hidden,  kennedy2014internal}.

On the other hand, non-representational significance attributions are of the type that are immediate (as opposed to deliberative) and of a phenomenal character.  
They distinctly mean something to the experiencing subject. Examples include moments of intuitive insights, profound personal revelations, states of transcendence  \cite{rosenblatt1994intuition, mcdonald2008nature,  baruvss2017transcendent } (and more such examples discussed in Table \ref{tab1} below).  While the phenomenology of these experiences can certainly  be  investigated qualitatively using introspective methods, the full scope of the meaning experienced in all of these instances is only available through  first-person perspective.  \\  \\  {\sf  For these reasons, such subjectively attributed meaning is an observer-dependent construct: it literally holds in the mind of the beholder!}  \\   \\  Meaning of this form is a genuinely subjective quality\footnote{As such, the "objective" use of meaning in language and social communication is only a matter of standardization of what symbols may refer to, or formation of social norms and conventions within and between groups of agents   \cite{ray1968language, lewis2008convention, freire2020modeling,  arsiwalla2023cognitive  }. }. In the discussion that follows, we will elaborate  how  this  form of meaning might be directly tied to the qualia of consciousness, and how it may in fact manifest as the phenomenal content of all experience.  We also point to phenomenological evidence in favor of this view. 

\begin{remark}
\label{remimage}
Earlier, we mentioned about meaning as discussed in the sense of  Brentano's notion of  "intentionality"   \cite{ brentanopsychology }.  There, meaning is conceived as a relation between the physical and the mental  (see also  \cite{atmanspacher2022dual, atmanspacher2024psychophysical  }).  Later, in section 4, we will indeed formalize such a relation between the physical and mental domains. How does this relational notion of meaning corroborate with  the idea of subjective meaning being proposed here?
 
Notice that, meaning, when conceptualized as a relation between the physical and the mental, is necessarily an extrinsically defined notion.  Namely,  extrinsic to any physical and mental observer. On the other hand, one may ask, from the intrinsic point of view of an observer within the mental domain, how does this physical to mental relation manifest? In other words, what does the mind experience of this relation? If one thinks of this relation as a formal map from a domain (physical) to a codomain (mental), then the answer is: the image of this map within the codomain. Subjective meaning is thus the mental image of Brentano's intentionality. This is the facet of meaning that a conscious agent   can experience with their mind. For this reason, we could have also referred to this notion of meaning as intrinsic meaning.  The full relational map is only accessible to a metaphysical observer.  Working from such an intrinsic point of view is of direct relevance to any experience-based theory of consciousness.      
\end{remark}

\begin{remark}
Let us make a brief comment about qualia in more general modalities. Even though a sizable fraction  of  works discussing qualia often anchor around sensory experiences ("the redness of red", for instance), it is to be noted that qualia also refer to aspects of conscious experiences beyond the purely sensory   \cite{sep-qualia, merleau2013phenomenology, chalmers1997conscious}. Examples include; (a) qualia associated to subjective feelings and emotions, like the feeling of joy, sadness, or fear, which might not be directly tied (though indirectly related) to sensory input;    (b) qualia associated to thought processes, such as the immediate awareness of having a thought, a mental image or an idea; (c) qualia  associated to introspective states, such as the subjective sense of self-awareness, reflective thought, or experience of one's own mental states; to name a few (see \cite{sep-qualia, merleau2013phenomenology, chalmers1997conscious} and references therein).  The above and other examples of qualia not  exclusively tied to sensory perceptions will be relevant for us later.   
\end{remark}

Empirically, how could subjective meaning  manifest as the phenomenal content  of qualia (in the broader sense of qualia mentioned in the remark above)?  For instance, consider the case of qualia associated to emotions: one may ask whether emotional feelings might come endowed with some form of subjective meaning that is immediate to the experience?  Now, emotions and feelings as intricate mental states certainly do carry a personal form  of significance for the experiencing subject.  Furthermore, this form of significance is highly subjective, contextual to the individual's history, and intimately tied to their    internal states.  But significance of this kind is in fact  a semantic quality, in that, it means something to the subject  (in the next   section we compare this notion of subjective meaning to Frege's notion of sense).  It is this aspect of meaning that we claim is raw and  immediate to the emotion (as opposed to a form of deliberative interpretation of the emotion).   Now,  discussions about phenomenal properties usually refer to the immediate, intrinsic and non-representational  qualities of conscious experiences that constitute the "what it is like" aspect of those experiences    \cite{nagel1974like}.  The form of meaning that we are referring to above, is intrinsically subjective  (involving  personal significance and value attribution)  and non-representational in the same way that sensory qualia are. One may report one's emotional experiences categorically, but there is no way to convey an objectively third-person perspective as to what that emotion means to the experiencing subject. This meaning or significance  is closely associated to the raw feel of the emotion itself. 

First, let us discuss why meaning of the above kind may be closely entwined with the subject's own self-states.

\begin{remark}
Let us note that issues related to personal significance or personal value attribution within the realm of a subject's experiences are features closely associated to their self-model.
\end{remark}

\noindent Therefore: 

\noindent\fbox{%
    \parbox{\textwidth}{%

\begin{proposition}
 We posit that subjective meaning is phenomenologically relevant to all  qualia associated with the subject's self-states, or at least those involving any relation between self-states to other states within the subject's  experiential realm.   
\end{proposition}

    }%
}
\vspace{.3cm}

From a systems theory perspective, a self-model is a construct that encompasses an individual's  interpretations, value attributions and  representations concerning their own identity, internal states, belief systems, emotional states and behaviors  \cite{metzinger2004being, metzinger2007self}. This internal model contributes to the recognition of one's own goals and value systems, and consequently,  it   influences all external models of the subject (such as those pertaining to the world, or models of other agents that the subject interacts with). In the more specific context of consciousness studies, the relevant kind of self-models are what are referred to as "phenomenal self-models". These models  were  proposed  by Thomas Metzinger  in 2004  \cite{metzinger2004being}. They are internal dynamical representations of the agent, and carry phenomenal content. A feature of these self-models is that they not only constitute a representation of the agent, but they  also  co-represent the representational relation (between the agent and the world) itself  \cite{metzinger2004being, metzinger2007self}.  In other words, they are equipped with a phenomenal model of the intentionality relation\footnote{According to Metzinger, a system of "third-order embodiment" possess phenomenal self-models \cite{metzinger2007self}. Such a system explicitly models itself as an embodied being, and maps some of the representational content generated in this process directly onto conscious experience. In other words, it consciously experiences itself as embodied. }.   
According to Metzinger,  phenomenal self-models are characterized by   phenomenal properties pertaining to ownership,  perspectivalness and selfhood; all of which are essential for creating a first-person perspective \cite{metzinger2004being, metzinger2007self}.  Presumably,  these phenomenal properties realize what is referred to as sense of ownership, sense of situatedness  and sense of self respectively.  The "sense" in the above terms is indeed a form of subjective significance or knowingness, and hence a form of meaning (the link between sense and meaning is further discussed in the next section).  For that reason, an experience of self, necessarily features subjective meaning of the kind we have noted above. 

Furthermore, one may extrapolate that the above stated properties of  a  phenomenal self-model would also ascribe subjective significance to the very manner in which an agent interacts with and perceives the world. In other words, the subjective meaning attached to (or in part, defining) one's self attributes, will influence the way a subject experiences the world from their first-person perspective. Their meaning-attributes of self affect their meaning-attributes of the world (and vice-versa, owing to feedback from the environment). 



Let us now discuss literature examples which demonstrate the above.  What we are suggesting here that subjective meaning,  as the phenomenal content of experience, may well be ubiquitous to consciousness.  Table \ref{tab1}   summarizes various examples of conscious  experiences, typically those  besides the purely sensory ones, which explicitly evoke a personal form of meaning for the experiencing subject. The text thereafter, further elaborates  on these.   

\begin{table}[h!]
\centering
 \begin{tabular}{|| c | c ||} 
 \hline   
1. &  Spiritual realization and personal revelations  \\  [0.5ex]   \hline  
2. & Eureka moments and intuitive insights  \\ [0.5ex]  \hline  
3. & Experiences of aesthetic appreciation  \\  [0.5ex]   \hline  
4. & Feelings of love and empathy  \\  [0.5ex]   \hline  
5. & States of enlightenment and transcendence  \\  [0.5ex]   \hline  
6. & Altered states of consciousness  \\  [0.5ex]  \hline  
7. & Feelings of synchronicity   \\  [0.5ex]  \hline  
8. & Out-of-body experiences  \\  [0.5ex]  \hline  
9. & Near-death experiences  \\  [0.5ex]  \hline  
10. & States of flow  \\  [0.5ex]  
  \hline
 \end{tabular}
\caption{Examples of conscious experiences that constitute subjective meaning. }
\label{tab1} 
\end{table}

\begin{enumerate}
 
\item   Moments of personal revelation, existential realization or sudden transformative life experiences like a spiritual realization, a meditative revelation or an epiphany  have been discussed in  \cite{greenberg2004handbook,  mcdonald2008nature, miller2013oxford, austin1999zen, van2017brain}.  
These experiences have been reported to involve a deep and immediate  personal significance to the experiencing subject. For that reason, the aforementioned studies investigating these experiences lean towards consciousness, rather than seeking explanations based on purely cognitive control or processing.  

\item  Sudden realizations such as "eureka" moments, intuitive insights, instinctive or gut feelings, that provide an immediate sense or perceived understanding to complex problems also possess subjective meaning within conscious experiences (see \cite{rosenblatt1994intuition, volz2006neuroscience}).  Likewise, for introspective awareness; that is, the immediate awareness of one's own mental states, thoughts or introspective processes \cite{van2000inward}.

\item  Profound experiences of artistic or aesthetic appreciation, such as being deeply moved by a piece of music, art or even nature.  Feelings of awe, wonder or personal connection when experiencing aesthetic perceptions.  All of these  constitute examples of conscious experiences that invoke an immediate kind of   meaning in the form of personal significance  \cite{clarke2011music,   epstein2004consciousness, reinerman2013neurophenomenology }. 

\item  Intensely felt connections in personal relationships, profound moments of empathy or experiences of deep understanding and resonance with others, including feelings of love, are examples of experiences deeply tied to a purely subjective notion of meaning  \cite{thompson2001empathy,  zeki2007neurobiology,  cacioppo2012social }.

\item Moments of transcendence (either through meditation, contemplation or religious practices),  perceived states of enlightenment, metaphysical awareness, spiritual awakening, mystical encounters eliciting  immediate, ineffable feelings of unity and connectedness either with a higher state of being or a divine presence \cite{maslow1969various, hawkins2015transcending, baruvss2017transcendent, forman1998does, taves2020mystical  }.  These too, hold deep personal significance for the subject.  


\item  Experiences involving altered states of consciousness, induced by meditation, trance or altered perceptions; such as those during lucid dreaming, mindfulness, heightened self-awareness, or those induced by hallucinogens  \cite{timmermann2023neurophenomenological, bayne2018dimensions, milliere2018psychedelics }.


\item  Following Carl Jung,  there are several literature studies  involving  instances of synchronicity or meaningful coincidences that evoke immediate, ineffable feelings of  interconnectedness for the experiencing subject \cite{jung1985synchronicity, Peat1987, donati2004beyond, colman2011synchronicity }.

\item  Instances of transpersonal experiences, such as out-of-body experiences, which involve a sensation of being detached from one's physical body and perceiving the world from a location outside the physical body  \cite{blanke2004out, blanke2016leaving, metzinger2005out }.  These experiences are intimately concerned with sense of self. 

\item  Also, near-death experiences such as experiences of vivid and panoramic life reviews, where subjects relive or review significant events or memories from their lives in a  comprehensive manner  (these are sometimes reported with feelings of peace or transcendence while seeming to enter into a different realm during the experience)  \cite{blackmore1996near, van2011near, blanke2016leaving }

\item  Lastly,  experiences of being in a state of flow, where immediate, ineffable insights and intense focus arise during engaging activities like creating art, playing music or solving complex problems  \cite{jackson1996development, nakamura2002concept, wrigley2013experience }.

\end{enumerate}

The above are all examples of conscious experiences (although not of a typically sensory type) that  constitute some  form of personal meaning for the experiencing subject. We claim that the  phenomenal content of these experiences is of a semantic type, that is, it is subjective meaning.

\section{Do Raw Feels Have Subjective Meaning? }

The experiential phenomena enumerated in the section above, while predominantly  of a non-sensory nature, have nonetheless been extensively  studied in the consciousness literature (see references cited above). The important point of the above discussion was that a personal notion of meaning is ubiquitous to a wide variety of conscious experiences  (particularly, those referring to non-sensory modalities). What about the raw feels of consciousness? Are they associated to meaning? 

Expositions concerning sensory qualia are often discussed in terms of raw feels    \cite{sep-qualia}.  In order to make the case that raw feels too, cannot be devoid of meaning that is intrinsic to the experiencing subject, let us  consider a bridging concept between feels and meaning: namely, that of "sense".  In philosophy of mind, the term sense has multiple connotations: it may refer to a subjective perception, an understanding or significance of something, or a form of knowledge  \cite{frege1892sinn, austin1962sense, austin1979other }.  Sense, perfectly well, applies  also to experiences beyond those involving the purely sensory modalities (for instance, those involving internal states).  Here we will anchor our discussion mainly on Frege's use of the notion of sense  \cite{frege1892sinn}. 

\begin{remark}
\label{rem3.1}
In his seminal 1892 paper, "On Sense and Reference", Gottlob Frege  developed his ideas on sense and reference  \cite{frege1892sinn}.  This laid the groundwork for the metaphysics of thought, theories of meaning in philosophy of mind, as well as discussions on semantics and  language
\footnote{Following Frege's seminal essay, other works that have extensively addressed the relation between language, meaning and mind include those of Bertrand Russell,    Ludwig Wittgenstein, Hilary Putnam and Saul Kripke,  to name a few   \cite{russell1905denoting, ludwig1953philosophical, Putnam1975, kripke1980naming }. }.    Frege elaborates on the distinction between sense and reference, and suggests that both, sense and reference contribute to  meaning  (with respect to both, expressions as well as thoughts).  

According to Frege, reference\footnote{Some authors also use the term "denotation" for "reference" \cite{sep-frege}. }  is concerned with naming objects or entities. It refers to, or is about, something.  Sense, on the other hand,  characterizes the mode in which something presents itself. In terms of expressions, this mode  accounts for what the expression signifies (to the subject).   It is the way by which one understands the expression via internal conceptual relations.  Given the reference of words (as names that refer to something), the reference of a sentence as a whole can be described in terms of the references  of the individual words and the way in which those words are arranged in the sentence. Based on this  Frege identifies  the reference of a sentence  as a truth value  (see also  \cite{sep-frege,  Dummett1983} for  an interpretation of Frege to the philosophy of language). Correspondingly, the sense of a sentence is a "thought"  (\cite{frege1956thought}  discusses the metaphysics of thought, as well as its linguistic realization).  While the reference of a sentence can assume one of two truth values, the sense of a sentence can be acquired from a potentially infinite  possibility of thoughts. Thus, sense and reference are both, compositional and hierarchical, existing at each level of the structure of the expression. Two syntactically distinct expressions, may have identical references (at each corresponding level of their structures), but distinct  senses. 
\end{remark}

The relevant question then is: Do raw feels have a subjective "sense" for that what is being experienced?  Is this a form of meaning, after all?  Let us now answer these questions. In doing so, we will make use of the following proposition:  

\noindent\fbox{%
    \parbox{\textwidth}{%

\begin{proposition}
Mental representations are to the mind, what symbolic expressions are to language.  
\end{proposition}

    }%
}
\vspace{.25cm}


Building up on remark \ref{rem3.1} above, let us analyze Frege's notion of sense in the context of mind. In Frege's view, the "sense of an expression" is a semantic attribute based on personal significance to the subject. This sense conveys a meaning of the expression to the subject. In this context, Frege even described thought as a semantic construct.  In order to systematically lift  this notion from the philosophy of language to the philosophy of mind; first, notice that just as an expression is a syntactic structure based on rules of composition and an architectural hierarchy; so is a mental representation of objects in the world (or of internal states),  a syntactic construct  based on rules of composition and structural hierarchy\footnote{Psychophysical studies involving preverbal human infants have demonstrated compelling evidence of the use of logical reasoning and probabilistic inference in the way humans learn and construct  mental  representations of objects in the world \cite{teglas2011pure,  cesana2018precursors}. These early representations are constructed using logical primitives and rules of composition \cite{piantadosi2016logical}.  }  (in the following section, we elaborate further on how grammar extends beyond spoken language; see also  \cite{zhu2007stochastic, mumfordblog }).  Hence, our proposition above: that mental representations are to the mind, what symbolic expressions are to language. By themselves, they are both syntactic constructs based on admissible structural compositions of linguistic and mental primitives respectively. Extrapolating Frege's philosophy to all syntactic constructs, mental representations also have a reference and a sense. Their  reference is literally what they are about, or what they represent; whereas, the sense of a mental representation is simply its significance to the subject. The latter is thus a semantic attribute of the representation. Mental representations thus have a structural (syntactic) aspect, as well as a semantic aspect. In the phenomenology of consciousness, when one speaks about awareness, or being aware about "something", it is certainly not the "thing" itself\footnote{See \cite{doring2010thing} for discussions on mathematical foundations of "what is a thing?"; and \cite{zalta2012abstract}  for a discussion on the metaphysics of abstract and concrete objects. }  that the subject is, or even possibly could be aware of\footnote{Several philosophers including Husserl  and  Merleau-Ponty  have emphasized this point  \cite{husserl1913ideas,  merleau2013phenomenology } .  }.    Furthermore, the fact that qualia are transparent, implies that one does not experience their mental representations as representations either.  Rather, it is only certain attributes of one's mental representation (of the  "thing" being "observed")  that the subject gains awareness of.  And, a prominent attribute of a representation is its subjective  meaning.  \\   
 \\  
{\sf Based on the above, we surmise  that it is subjective meaning of one's mental representation of the "thing" being observed (either external or internal) that the subject is aware of during a conscious experience. The raw feel of a conscious  experience, therefore, has a sense to it, coming from the mental representation of that what is being experienced. }

\noindent  Two remarks are in order here: 

\begin{remark}
There is of course room to argue that there may be other attributes of mental representations such as its symbolic, computational or   informational features that also partake in the realm of awareness. However,  unlike  meaning, these other features are not non-representational. Which is why, they do in fact admit a third-person description. However, qualia, having a phenomenal character, are deemed  non-representational. For that reason, we maintain that subjective meaning is the more compelling candidate that captures the phenomenal character of qualia.    
\end{remark}

\begin{remark}
Let us also briefly comment on empirical studies that also invoke the notion of sense in consciousness research. In empirical and clinical literature on consciousness, one often encounters sense via terms such as,   "sense of agency",  "sense of self", "sense of ownership",  "sense of time", and so on  \cite{de2004sense, hohwy2007sense, martin1995bodily,  wittmann2013inner, lopez2021sense  }.   In all of these contexts,  sense is often described as  a subjective perception or comprehension of something. Notice that such a description is precisely a semantic attribute based on personal significance to the experiencing subject.  The same can be said for other examples of sense such as an intense sensation of pain, an abrupt sense of discomfort, or a sudden surge of pleasure.  In fact, sense and feeling are sometimes used interchangeably in this literature. 
\end{remark}




\section{The Formal Structure of Meaning}

So far, we have put forth, mostly on phenomenological grounds, that meaning,  generated intrinsically by the experiencing self, is an integral and inseparable characteristic of qualia, across the board. Furthermore, the subjectivity, non-representational and as well,  the transparency of, what has been referred to as subjective meaning, suggests why it may be directly linked to the phenomenal contents  of conscious experience. However, to simply label meaning as a mental property, does not do full justice to the important role that meaning plays with regard to the mind-matter relation. This view has also been proposed in \cite{atmanspacher2022dual, atmanspacher2024psychophysical }, albeit, based on different motivations. To elucidate the character of meaning with respect to the nature of qualia, and also its relevance to the mind-matter relation, we first need to discuss the formal structure of meaning itself. 

Of course, the concept of meaning is not exclusive to the philosophy of mind. Rather, it has a shared history across disciplines, including  linguistics, semiotics, programming language theory, type theory, category theory and  metamathematics, to name a few. Based on the specific domain, one considers or constructs  appropriate semantic frameworks for given syntactic systems (both, verbal or non-verbal). A semantic framework provides an interpretation space for syntactic expressions.  For instance, a symbolic expression such as a sentence or a logical proof, by itself, is a purely syntactic construct. It is its semantics that elucidates what the sentence or proof refers to or how it can be interpreted.  For this reason, one works with  a formal  interpretation space within which syntactic expressions may be evaluated. These evaluations may either yield truth values or, more generally, meaning vectors (see distributional models of meaning  \cite{coecke2010mathematical}).  Semantic models thus enable value or categorical evaluations of the syntax.  However, what is important are not the specific categories or values that a syntactic structure may acquire within a model   (which often depend on model constraints); rather, it is that these evaluations establish  relations between syntactic expressions. For example, relations of similarity. It is such relations that are hypothesized to give words and sentences their meanings, which is how the so-called   "compositional distributional models of meaning"  operate  \cite{clark2008compositional}.

Let us consider the case of natural language semantics. Here, one may want to study lexical semantics which refers to word meanings, and how meanings between similar words might relate. In that case, an often used dictum due to J. R. Firth goes as, "You shall know a word by the company it keeps"  \cite{firth1957synopsis}.  That is, words with similar linguistic distributional properties (in large corpus  data) have similar meanings. This is known as the distributional hypothesis of meaning \cite{harris1954distributional}.  Additionally, besides word meanings, one may also want to investigate meanings of phrases and sentences (given the meanings of the words contained therein). In this case, one resorts to the "principle of semantic compositionality",  which is also known as  Frege's principle \cite{pelletier1994principle}.  According to this, the meaning of a complex expression is derived from the meanings of its constituent expressions and the syntactic rules used to combine them\footnote{Of course, compositionality is not the only guiding principle for semantics of complex expressions or conceptual combinations.  One also has to take into account contributions from non-compositional semantics, as illustrated  in  \cite{bruza2015probabilistic}.  These authors emphasize that the  latter is markedly distinct from the former in terms of  contextual dependence. This additional contribution is necessary for a more comprehensive treatment of natural language semantics and pragmatics. However, for our purposes here, compositional models will suffice for what follows. Additional contextual contributions can be included thereafter. }.   Hence, the  meaning of  a sentence is then obtained by putting together these two principles, which yields what is known as the "Compositional Distributional Semantics" framework   \cite{coecke2010mathematical}.  The mathematical formalization of this framework in terms of monoidal categories of vector spaces and Lambek's pregroup  grammar  enables one to compute the meaning of well-typed sentences from the meaning of its constituent words, by inducing  the type reduction mechanisms of the pregroup grammar to the whole category. These sentence meanings live in a single space, realized via monoidal (tensor) products of the category.  Meaning is thus evaluated via  a category-theoretic functor from a syntactic space governed by rules of grammar, to a semantic space, functioning as an interpretation space for syntactic constructs.  

Interestingly, such a functorial  perspective (as elucidated above) turns out to be much more general than its particular instantiation in linguistics alone.  In fact, in mathematical logic and abstract algebra this goes by the name of "Functorial Semantics"  \cite{lawvere1963functorial}.   For example,  an association map from a given syntactic system to a semantic system, such as:   
\vspace{.2cm}
\begin{eqnarray}
{\bf Association \,\, Map \, : \, Syntactic \,\, Space \, \to \, Semantic \,\, Space}
\label{mm}
\end{eqnarray}

\noindent may realize how abstract algebraic systems yield a geometric representation within an appropriate space (for instance, the representation theory of groups).  In this case, the "intuition" or physical interpretation of the algebraic system is a geometric one.  

\begin{remark}
\label{sem1}
It is interesting to see how eq.(\ref{mm}) compares to Charles S. Peirce's Theory of Signs, or Semiotics  \cite{peirce1883studies,  peirce1977semiotics}. Semiotics refers to a formal theory of  signification of entities (across science and philosophy), their representation, in terms of suitable signs, and the meaning of those signs  \cite{ sep-peirce-semiotics  }.  The essential elements of Peirce's theory include a triad consisting of a sign $S$, the object $O$, and the interpretant $I$. The sign can be thought of as a symbol or a signifier of the thing (the object) being signified. Whereas, the interpretant is the understanding (sometimes described as a "translation") that the user of the sign has of the object-sign relation  \cite{ savan1987introduction, sep-peirce-semiotics }.   

The object determines the sign, and in turn, the sign, via the specific representation of the object, influences and shapes the interpretant. The order of these relations is so:  
\begin{equation}
O \to S \to I    \nonumber
\end{equation}
One can think of the relation $O \to S$ as a $Representation$ of the object, while the relation $S \to I$ can be viewed as the $Interpretation$ of the sign (or an $Association$ that maps signs to what they refer to). The meaning of the sign, then manifests via the interpretation of the representation of the object being signified.  

The comparison to eq.(\ref{mm}) is now straightforward: the sign $S$ belongs to the syntactic domain, whereas the interpretant $I$ is semantic. The interpretation map $S \to I$, then corresponds to the association map in eq.(\ref{mm}).  
\end{remark}

Now, let us  take the above formal perspective of meaning, as a  map from syntax to semantics (eq.(\ref{mm})), more seriously. How does one apply this to the mind-matter problem?  To do so, we propose the following  maps  of relations between the physical and  mental domains:

\begin{equation}
\begin{tikzcd}[column sep=15em, row sep=10em]  
Obj_P  \arrow[d, "\scalebox{1.5}{${\cal R}_{Obj}$}" '  ]     \arrow[r, "\scalebox{1.5}{${\cal A}_P$}",  shift left=0 ]  &  Rel_P   \arrow[d,  "\scalebox{1.5}{${\cal R}_{Rel}$}" ]      \\ 
   Obj_M    \arrow[r, dashed, "\scalebox{1.5}{${\cal A}_M$}" ',  shift left=0 ]  &   Rel_M  
\end{tikzcd} 
\label{dia1}
\end{equation}

This diagram requires some unpacking. Let us first motivate the relations shown therein, and then examine the structure of meaning appearing here and how that relates to qualia, discussed above in earlier sections. After that, we indicate how this system of maps weighs on the mind-matter relation.

The top row maps collections of physical objects $Obj_P$ to relations between these, indicated by $Rel_P$, via the association map     
\begin{equation}
{\cal A}_P : Obj_P \to Rel_P    \nonumber
\end{equation}
 In fact, the association map ${\cal A}_P$ defines  relations contained within $Rel_P$. Note that these relations can be of general arity (in which case, the objects in $Rel_P$ may be diagrammatically presented as hyperedges, and the ${\cal A}_P$ denote multi-source / target maps for constructing hyperedges). Together,  the collection $\{ Obj_P,  \,  {\cal A}_P ( Obj_P )  \}$  forms the physical domain, denoted ${\cal P}$. 

Likewise, $Obj_M$ denotes collections of mental objects, and $Rel_M$ contains relations (unary, binary or any higher arity) between mental objects. Here again, the association map  
\begin{equation}
 {\cal A}_M : Obj_M \to Rel_M  \nonumber  
\end{equation} 
  defines  relations contained within  $Rel_M$. We will refer to the collection $\{ Obj_M,  \,  {\cal A}_M ( Obj_M )  \}$  as the mental domain ${\cal M}$. 

Furthermore, the two representation maps 
\begin{equation}
  {\cal R}_{Obj} : Obj_P  \to  Obj_M \qquad  \mbox{and}  \qquad  {\cal R}_{Rel} : Rel_P  \to Rel_M
\end{equation}  
   take objects and relations in the physical domain and construct models of these objects and relations respectively within the mental domain. At times, one interchangeably uses the term "mental representation" to refer to these models, rather than the maps themselves\footnote{More generally, mental representations may also encompass models of internal states, such as one's self model (see \cite{sep-mental-representation} for an overview). }. While mental models may well supervene upon the brain's neuronal circuitry (and feedback loops with the environment, through the body), strictly speaking, these models are directly accessible only to the cognitive / conscious agent. For that reason, these models lie solely within the mental domain\footnote{The neuroscientist experimenting upon the agent, is merely inferring what the model may be about. }, while the representations ${\cal R}_{Obj}$ and ${\cal R}_{Rel}$ themselves, are maps from the physical to the mental domain. 

Now, how does one realize meaning (in the sense of eq.(\ref{mm})) within the framework of eq.(\ref{dia1})? In order to answer this, first let us note that the  spaces of relations $Rel_P$ and $Rel_M$ above, precisely function as semantic spaces. In linguistics, and more specifically in natural language semantics, words (and sentences) that are deemed to carry similar meanings, are hypothesized to occur in similar linguistic distributions \cite{harris1954distributional}.  Semantic spaces are then constructed as vector spaces (or more generally, as metric spaces) endowed with a similarity metric, such that  words or sentences  with similar linguistic distributions are  clustered together, and thus exhibit similar meanings.  In other words, the meaning of a syntactic construct is a relational attribute that depends on how that construct is related  (with respect to the similarity metric) to other constructs within that space. Meanings of concepts can be extracted from its network of relations with other concepts.  Examples of such models include word embedding models in computational linguistics  \cite{mikolov2013distributed},  and   Gardenfors'  "conceptual spaces"  in cognitive science  \cite{gardenfors2014geometry}.    
Furthermore, there is compelling evidence from neurophysiological studies demonstrating how semantic maps are encoded  in the human brain as a network of (encoded) concepts,  via similarity distance in encoding space   \cite{huth2012continuous, popham2021visual,  pacheco2021volitional }.  Despite different choices in implementation, all the examples above suggest that relational maps between concepts provide the building blocks of semantic spaces.

What about the space of syntax? Where does that appear within our   framework?  Syntactic structures refer to admissible compositions of symbols of a formal language, that generate the words or sentences of that language. What determines admissibility of symbolic compositions is the formal grammar specifying the language. In recent years, what one means by language and grammar has vastly generalized beyond traditional forms of written language. Examples include graph grammars in graphical languages  \cite{marcolli2015graph,  arsiwalla2023pregeometry, arsiwalla2021pregeometric,  arsiwalla2022homotopies  },  pregroup grammars in diagrammatic languages involving string diagrams \cite{lambek2008word, coecke2010mathematical },  visual grammars corresponding to compositions involving visual scenes and actions  \cite{zhu2007stochastic, mumfordblog }, and hypergraph / operator algebras that extend graph grammars \cite{ arsiwalla2023operator,  zapata2023hypermatrix,  zapata2022invitation  }, 
just to name a few.  
In our case here, the collections of objects in both, $Obj_P$  and  $Obj_M$  are not arbitrary collections. How objects interact and compose in the physical world, and the ensuing physical properties of these compositions, is highly constrained by the laws of physics; including laws of dynamical evolution,  order effects such as causal ordering, kinematical constraints such as conservation laws, and  geometrical constraints due to the spaces within which these objects live. Furthermore, hierarchical structures of physical systems             have their own macro-level laws in the form of statistical mechanics and thermodynamics.  Consequently, representations of the physical world in $Obj_M$ are also modeled accordingly. The visual grammar discussed in \cite{zhu2007stochastic },  the "physics engine" in \cite{lake2017building}, and the "impossible figures" in  \cite{macpherson2010impossible}, all  lend strong evidence towards a grammar underlying mental  representations\footnote{Needless to say, there have been numerous debates in the literature as to whether grammar, in some form, may be innate to cognitive agents  \cite{gopnik1997inheritance}. }.   In other words, structures that "live" in $Obj_P$  and  $Obj_M$, realize syntactically well-defined structures following a broader notion of grammar, as indicated in the examples and references above. 

Therefore, on one hand, we have rules constraining the space $Obj_P$ (and its representation in $Obj_M$), that allow for only very specific compositions of physical (respectively, mental) objects corresponding to existing constructs in the world (mind). On the other hand, we have an association map ${\cal A}_P$, that defines relations between physically permissible objects, leading to the space of physical relations $Rel_P$, which in turn, functions as a semantic space for the physical domain. Correspondingly, we have the association map ${\cal A}_M$, defining relations between permissible mental  objects (obtained as representations of physical objects or of internal states of the agent), and leading to the space of mental relations $Rel_M$, which now  functions as a semantic space for the mental domain. As per eq.(\ref{mm}), in the system in eq.(\ref{dia1}) we now have two possible levels at which meaning may manifest as a distributional and compositional attribute of these mappings from syntactic to semantic spaces: the physical level, as well as the mental level. However, these levels are not independent, and neither is the kind of meaning associated to these domains.

Our central question is  how does subjectively attributed meaning arise in the mind, and how does that associate to the broader mind-matter relation?  Our map ${\cal A}_P$  suggests that a formal notion of meaning also ought to exist within the physical domain  (in terms of relations and interactions between physical structures, constrained by causal and statistical laws).  However, there needs to be something to read-off this meaning associated to the physical world\footnote{One may well call such an entity an "Observer". Though this is by no means a sufficient condition to define an observer.  Hence, to read-off any kind of meaning, an observer is imperative.  A discussion about observer theory from the perspective of second-order cybernetics can be found in \cite{ elshatlawy2023ruliology }.  }.  Now this is where mental states come into the picture. They serve as a means to read-off associations in the world. This is how that may be so: through the representation maps ${\cal R}_{Obj}$ and ${\cal R}_{Rel}$, models of physical objects and their relations are constructed within the mental domain. Then, following simply the commutativity of the diagram in eq.(\ref{dia1}), the association map  ${\cal A}_P$  (within the physical domain)  induces the association map ${\cal A}_M$ (within the mental domain).  Thus, meanings about things in the physical world can be read-off via mental states, following this  relation between the physical and mental domains.

\begin{remark}
Once again, notice the parallels to Peirce's object-sign-interpretant triad in semiotics. The $O \to S \to I$ relations in Remark \ref{sem1}  are precisely captured by the following sequence of maps in eq.(\ref{dia1}): 
\begin{equation}
Obj_P  \to  Obj_M  \to  Rel_M    \nonumber 
\end{equation}
where elements of $Obj_M$ play the role of signs with respect to the mind-matter relation, and $Rel_M$ give rise to Peirce's interpretant, corresponding to mental objects. Of course, the above is only a particular application of Peirce's theory of signs, which otherwise, was intended for broader applications across the sciences. 
\end{remark}

Returning to our discussion about the meaning of things in the physical world,  the astute reader will immediately notice that this is not the only instantiation of meaning available to one's mental faculties. That is indeed so. And this is where internal states, as well as perceptual limitations of the subject, matter.  Let us discuss this now:  
 
\noindent (i) First, as has been well documented in the literature, mental representations of objects and relations in the physical world, need not always be complete or accurate descriptions of those physical entities \cite{sep-mental-representation }. In practice, the cognitive / conscious agent has to build models of both, the world as well as that of other agents  from incomplete    social, behavioral and sensorimotor data  (either due to stimulus uncertainty, hidden states of other agents, or  the agent's own limitations in processing and parsing all that information),  about objects and agents in its environment,  obtained in real-time  \cite{conscious12016, arsiwalla2023morphospace,  arsiwalla2016three,  arsiwalla2017consciousness,   freire2023modeling, sanchez2017social,  freire2018modelingtom, demirel2021distinguishing }.

\noindent (ii) Second, and more importantly, minds can also model their own internal states. The self-model being a classic example. And these internal models within the mental domain  also influence each other. Again, the example being how the self-model modulates the way the agent probes its environment to construct  a world model, and how through its self-model, the agent perceives its situated and embodied self, immersed within the world. What this means is that there is a self-referential aspect within the design of mental models, which is arguably a peculiar feature of  the mental domain.   This self-referential feature, or circular feedback,  can  be modeled  via self-loops within the mental domain. This gives us  the following maps\footnote{Note that these are not identity morphisms, but non-trivial mappings within the mental domain. }:  

\begin{equation}
{\cal S}_{Obj} : Obj_M  \to  Obj_M  \qquad \mbox{and} \qquad {\cal S}_{Rel} : Rel_M  \to Rel_M
  \nonumber 
\end{equation}

\noindent  which extend the diagram in eq.(\ref{dia1}) by:
\vspace{.25cm}
\begin{figure}[h]
\centering
\begin{equation}  
\includegraphics[width=0.65\linewidth]{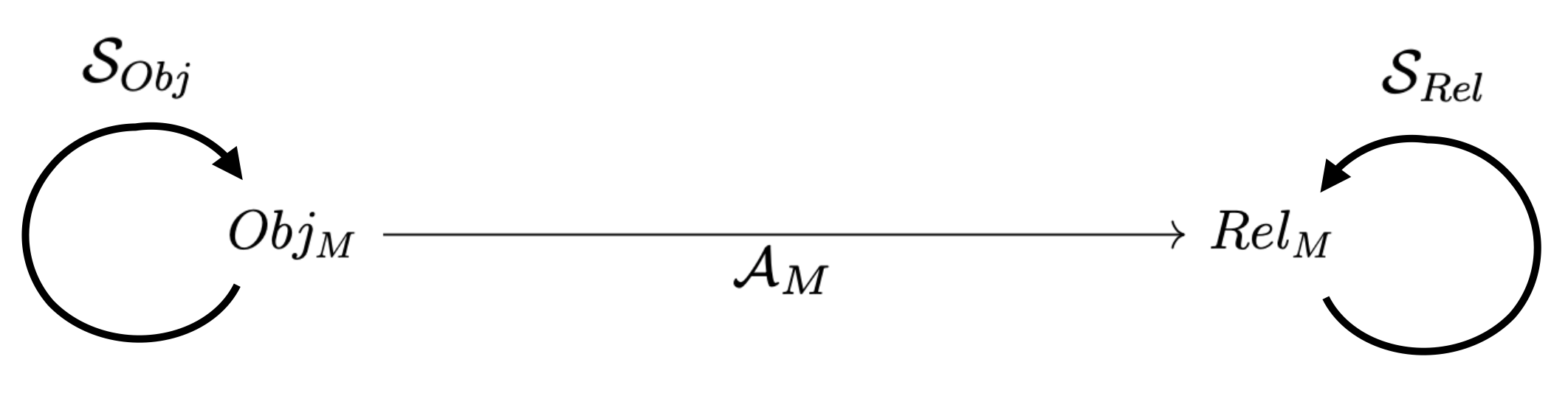}
\label{dia2}
\end{equation}
\end{figure}

In other words, in $Obj_M$ we have additional syntactic rules (compared to those in the purely physical domain) governing how mental models interact with each other. For instance, how the self-model (which includes models of the agent's psychological as well as bodily faculties)  may interact with or influence parameters within the subject's world-model in order to generate anticipatory and adaptive control responses to a planned action  \cite{metzinger2007self,  herreros2016forward, arsiwalla2019beyond, herreros2017cerebellar  }.  Correspondingly, the semantic space  $Rel_M$ also captures relations within and between all internal models.

\begin{remark}
The maps, in eqns.(\ref{dia1}) and (\ref{dia2}) above,  demonstrate  how  meaning involves relations between the physical and mental domains. The 
  mind-matter relation thus plays a central role in realizing subjective meaning within mental states.   How does one realize sense and reference within this framework?

The map from $Obj_P$ to $Rel_M$ following the composition\footnote{In mathematical literature such a  composition of maps is often written in right-to-left notation, but here we prefer the left-to-right presentation for the sake of better readibility.}  ${\cal A}_P \circ  {\cal R}_{Rel}$ suggests how reference (as a relation between the physical and the mental)  about things in the world, as perceived by an agent with mental states, may be formalized:-  As a map from structure in the world to relations in the mind, that factors through physical relations actually found in the physical world, and hence relies on the association map ${\cal A}_P$.   

On the other hand, the composition  ${\cal R}_{Obj}  \circ  {\cal A}_M$ suggests how sense (as a relation between the physical and the mental),  as perceived by the observing agent, may be formalized:-  As a map from structure in the world to relations in the mind, that now factors through  various mental models that the agent may have, and hence relies on the association map ${\cal A}_M$. However, to fully account for sense as this composition of maps, one has to admit a slight generalization of the diagram in  eq.(\ref{dia1}) such that it is commutative up to higher homotopy (in this case, a 2-morphism)\footnote{ We thank Giulio Katis for bringing this point to our attention. A thorough category-theoretic treatment of this diagram along with admissible higher morphisms will be reported in a forthcoming article. }.

As an example, consider a physical object; say, the planet Venus. Its reference map passes through physical relations in the world (such as the planet between Mercury and Earth) which constrain its concept within the mental domain; and that yields its meaning as reference. On the other hand, its sense map passes through mental objects (which are mental models that the mind uses to know about these objects). These mental models are based on how the subject has experienced the planet Venus, either as the morning star, or the evening star.   Hence with respect to sense as a relation, there are more than one possible paths that could be followed; which explains why there can be different senses for the same object.  

\end{remark}

\begin{remark}
Based on the above,  sense and reference are both  crucial  to a subject's  experience.  Major dissociations or disruptions between pathways related to sense and those to reference, would arguably be implicated in aberrant psychological conditions. 

For example, sense without reference would suggest domination of internal biases over external evidence, as in cases of hallucination or delusion. 
Reference without sense corresponds to a state of being unable to contextualize what is being observed. Therefore, both sense and reference play a vital role in the phenomenology of consciousness.  

Also, in future technology aiming to achieve Artificial General Intelligence (AGI), we posit that such a corroboration between sense and reference would, at the very least, be necessary (if not sufficient).   
\end{remark}

\section{Discussion } 

This work posits that subjectively attributed meaning is ubiquitous in conscious experiences. Our hypothesis is that subjective meaning (or, one may also refer to it as intrinsic meaning) is intrinsic to the qualia of conscious experience, in that, the phenomenal content of conscious experience is precisely this type of meaning. In other words, we claim that the content of conscious experience is semantic.  To support this proposal, we have discussed literature examples from the phenomenology of consciousness, which demonstrate the close   connection between meaning and experience. 

Our notion of subjective meaning closely relates to what Frege has referred to as "sense".  Furthermore, we have noted out how subjective meaning also parallels  Peirce's "interpretant" in semiotics.  As a special case of Peirce's theory of signs, Peirce's object-sign-interpretant triad is captured following the sequence of maps $Obj_P  \to  Obj_M  \to  Rel_M$, in eq.(\ref{dia1}).  

Building upon the above, we have argued that  raw feels can not be devoid of   subjective meaning. It is this form of meaning of one's mental representation (of the stimuli) that the subject is aware of during a conscious experience. The raw feel of a conscious experience has a sense to it (\`{a} $la$ Frege), one that comes from the mental representation of that what is being experienced.  

Furthermore, we have presented a  formal structure of maps that elucidates the role of subjective meaning in the context of the mind-matter relation. Exploiting the mapping between the physical and mental domains, we show how syntactic and semantic structures  are realized within both, the physical and the mental domain. In this sense, grammar and meaning transcend conventional written language.  Meaning is  realized as a relational attribute arising from a map that interprets syntactic structures of a formal system within an appropriate  semantic space. Based on this, we have provided an explanation  of how sense and reference may be formally realized within the mind-matter relation.  

Our notion of subjective meaning is the mental image of Brentano's intentionality.  It is this subjective facet of meaning that a conscious agent  experiences with their mind. For this reason, we could have also referred to this notion of meaning as intrinsic meaning. The full relational map of meaning (linking the physical to the mental) is only accessible to a metaphysical observer. Working from such an intrinsic point of view is of direct relevance to any experience-based theory of consciousness\footnote{Of course, one may certainly choose to consider other points of view, other than the one that is intrinsic to experience, as considered here. Each such case will have its own metaphysical considerations to take into account, as well as how that may translate to experience.  }. 

Given that this work places meaning of an intrinsic kind at the heart of the problem of consciousness, or the qualia of it thereof, it is reasonable to ask whether or not conscious awareness itself is a guise of subjective meaning, as structurally defined in eq.(\ref{dia1}) above? Though this view may appear to be in  contrast to other proposals, such as consciousness being integrated information \cite{tononi2012integrated }, or that, consciousness being  identical to a specific maximally irreducible causal structure     \cite{oizumi2014phenomenology },  or that, consciousness  being a form of    computation  \cite{dehaene2017consciousness };  it may well be that these are different pieces of the larger puzzle, and perhaps one may find convergence of some of these different ideas in future theories of consciousness. For instance, a suggestion of how Integrated Information Theory itself may be extended to include compositional meaning was already proposed\footnote{Presented at the Models of Consciousness Conference 2019, University of Oxford:  \url{https://podcasts.ox.ac.uk/xerxes-arsiwalla-computing-meaning-conceptual-structures-integrated-information-theory} }  in \cite{oxtalk}.  However, note that any such synthesis of ideas does not  imply that information, causality, computation or meaning can simply be used interchangeably as a theoretical basis for consciousness science. Each of these are conceptually distinct (and also differ in the way they could be used as a basis of a scientific theory).   However,  different ways of  looking at  the problem (of consciousness) may certainly be a very fruitful exercise. What this work attempts to highlight is that meaning, of the kind that is subjectively attributed, seems to account for the phenomenal and non-representational aspects of conscious experience. 

Taking Frege's philosophy of language and mind seriously, we surmise  that  that syntax and semantics are also relevant to a theory of consciousness.  In particular, mental representations are to the mind, what symbolic expressions are to language. Within such a philosophical framing, when one speaks about awareness about "something", one may then ask: precisely what attribute of the "thing" is it, that the subject is actually aware of?  The answer, subjective meaning  "of"  the subject's mental representation of the "thing",  seems better suited than alternatives  such as information, causality or computation.  That is not to say that information, causality or computation are not useful to study how consciousness may work. In fact, these are absolutely indispensable for quantifying neuronal processes in brain networks (or for that matter, artificial neural networks), that may correlate to empirical states of consciousness \cite{ arsiwalla2018temporal, arsiwalla2013iit, arsiwalla2016computing,  xda2018phy,  arsiwalla2016high, xda2016global,  arsiwalla2017spectral,  arsiwalla2017brain,  arsiwalla2018measuring, sarasso2021consciousness  }.  However, if one wants to explain what may constitute the non-representational  content of  phenomenal experience, then, the proposal of  subjective meaning seems to offer a promising new avenue for conceptualizing theories of  consciousness. Such a notion of meaning is the attribute (of the thing observed) that the subject may well be aware of, or may sense (\`{a} $la$ Frege)  during  conscious experience.  This hypothesis, of subjective meaning as the phenomenal constituent of qualia of conscious experience, also  suggests that one take ideas from the philosophy of language and semiotics  more seriously towards the advancement of a comprehensive theoretical framework that satisfactorily addresses the phenomenal character of  consciousness.  


\section*{Acknowledgments}

The author would like to gratefully acknowledge Harald Atmanspacher,  Dean Rickles and  Giulio Katis  for discussions and constructive feedback on this manuscript.

\bibliographystyle{splncs03}
\bibliography{test2}

\end{document}